\documentclass[pra,nobibnotes,reprint,secnumarabic,superscriptaddress,floatfix,showpacs]{revtex4-1}
\usepackage{amsmath}
\usepackage{graphicx}
\usepackage{dcolumn}
\usepackage{bm}
\usepackage{cases}
\usepackage{amsthm}
\usepackage{amsfonts}
\usepackage{amssymb}
\usepackage{graphics}
\usepackage{amsmath}
\usepackage{times}
\usepackage{color}
\usepackage{ulem}
\usepackage{subfigure}

\begin{document}


\title{Regular and exceptional spectra of the two-qubit quantum Rabi model}

\author{Jie Peng}
 \email{pengjie145@163.com}
 \affiliation{Key Laboratory of Modern Acoustics and Department of Physics, Nanjing University, Nanjing 210093, China}
 \affiliation{Joint Center of Nuclear Science and Technology, Nanjing University, Nanjing 210093, China}

\author{Zhongzhou Ren}
 \email{zren@nju.edu.cn}
 \affiliation{Key Laboratory of Modern Acoustics and Department of Physics, Nanjing University, Nanjing 210093, China}
 \affiliation{Joint Center of Nuclear Science and Technology, Nanjing University, Nanjing 210093, China}
 \affiliation{Center of Theoretical Nuclear Physics, National  Laboratory of Heavy-Ion Accelerator, Lanzhou 730000, China}
 \affiliation{Kavli Institute for Theoretical Physics China, Beijing 100190, China}

\author{Daniel Braak}
\affiliation{EP VI and Center for Electronic Correlations and
 Magnetism, Institute of Physics, University of Augsburg, D-86135
 Augsburg, Germany}

\author{Guangjie Guo}
\affiliation{Department of Physics, Xingtai University, Xingtai 054001, China}

\author{Guoxing Ju}
\email{jugx@nju.edu.cn} \affiliation{Key Laboratory of Modern
Acoustics and Department of Physics, Nanjing University, Nanjing
210093, China}

\author{Xiaoyong Guo}
\affiliation{School of Science, Tianjin University of Science and Technology, Tianjin 300457, China}

\author{Xin Zhang}
 \affiliation{Key Laboratory of Modern Acoustics and Department of Physics, Nanjing University,
 Nanjing 210093, China}
 \affiliation{Joint Center of Nuclear Science and Technology,
Nanjing University, Nanjing 210093, China}

\date{\today}

\begin{abstract}
We have studied the two-qubit quantum Rabi model in the asymmetric
case and its generalizations with dipole and Heisenberg-type
qubit-qubit interactions. The solutions are obtained analytically
with eigenstates given in terms of the extended coherent states.
These models are relevant to the construction of ultrafast two-qubit
quantum gates and quantum state storage. For identical qubit-photon
couplings, a novel type of quasi-exact solution which exists for all
coupling values with constant eigenenergy is found, leading to level
crossings within the same parity subspace even for non-identical
qubits. In contrast to the quasi-exact states of the single-qubit
model, the condition for these exceptional eigenstates depends only
on a fine-tuning of the qubit level splittings but not on the
coupling to the photon field. This makes them excellent candidates
for direct experimental observation within circuit QED.
\end{abstract}


\maketitle

\section{Introduction}\label{1}

\noindent The quantum Rabi model \cite{jc} describes in a very
simple way the interaction between light and matter, the latter
being modeled by a single spin-$1/2$ particle. It has found wide
application in quantum optics \cite{guo,tw}, circuit
QED~\cite{sola,tk}, cavity QED~\cite{s,ac,ih} and quantum
information~\cite{aj}. Theoretically, various investigations have
been undertaken to solve it \cite{schweb,swa,trave,vv}. An
analytical solution was obtained in \cite{br}, using the Bargmann
space of entire functions to model the bosonic degree of freedom
\cite{barg}. At the same time, developments in the field of circuit
QED have reached the ultrastrong qubit-photon coupling region
\cite{so,so1}. Ensuing work studied other aspects of the full Rabi
Hamiltonian, such as real-time dynamics and dynamical correlation
functions \cite{bra1,bra2a}, new methods to rederive the solution of
the same model \cite{ly,qing,braak}, and several generalizations
\cite{trav,pj,2bite,pj1,bra2,dicke}.

An important generalization of the single-qubit case, the two-qubit
system with various interactions \cite{cpl} is the simplest model of
the universal quantum gate \cite{jj,cp} and has therefore
applications in quantum state storage and transfer \cite{cp1,jm}.
The general two-qubit Rabi model has to be described for values of
the coupling in the ultrastrong and deep strong regime where the
rotating-wave approximation breaks down \cite{so}, to be relevant to
the recent developments in quantum optics \cite{ag,jlf,zrhk} and
quantum information \cite{afe,rg}.

In this paper, we will give its solution for the case of discernible
qubits analytically, using a generalization of the method used in
\cite{br}, and also with extended coherent state method \cite{qing}.
At the same time, we consider some types of qubit-qubit interactions
based on this model, including dipole interaction \cite{suda,ggc},
XXX \cite{fa} and XYZ \cite{aa,gs} Heisenberg interactions. These
generalized models allow for additional control of the system and
may thus be of interest for applications. The eigenstates are
obtained in terms of extended coherent states or Fock states. These
expansions form the natural basis for numerical studies of the
real-time evolution, unbiased by the truncation procedure
\cite{bra1}. For identical qubit-photon couplings, there exists a
novel type of quasi-exact solution for all the coupling values $g$
with constant eigenenergy, and the condition for its existence just
depends on the qubit energy splittings if the qubit-qubit
interactions are not taken into consideration. In contrast to the
single-qubit case, these exceptional eigenstates have finite photon
number and may be easily accessible in experiments, giving them
possible application in quantum computation. A well-known example of
these states are the spin-singlet ``dark'' states \cite{rod-lara}
for identical qubits, corresponding to a decoupling of the singlet
sector. Remarkably, they exist also if the permutation symmetry is
partially broken and the qubits are strongly coupled to the
radiation field. If the qubits do not interact, we have found one
such state for non-identical qubits. For a special choice of
interaction, two exceptional
 states of the novel type are present in the spectrum.

Although all considered models possess a $\mathbb{Z}_2$-symmetry,
this discrete symmetry is not sufficient to render them integrable
because the dimension of the spin space (four) exceeds the number
(two) of irreducible representations of $\mathbb{Z}_2$, according to the labeling criterion for quantum integrability introduced in \cite{br}.

The paper is organized as follows. In Sec. \ref{s2}, we obtain the
regular spectrum and the exceptional solutions of the two-qubit
quantum Rabi model using Bargmann-space techniques and extended
coherent state method. The exceptional solutions are given in closed
form. In Sec. \ref{s3}, we generalize the two-qubit quantum Rabi
model to include the dipole, XXX, and XYZ Heisenberg interactions
and obtain their regular and exceptional solutions. Finally, we draw
some conclusions including perspectives for future work in Sec.
\ref{s4}.

\section{Solution of the two-qubit quantum Rabi model} \label{s2}
\subsection{Regular spectrum of the two-qubit quantum Rabi model} The
Hamiltonian of the asymmetric two-qubit quantum Rabi model reads
($\hbar=1$)
\begin{align}\label{gq}
H_{tq}=&\omega
a^{\dagger}a+g_{1}\sigma_{1x}(a+a^{\dagger})+g_{2}\sigma_{2x}(a+a^{\dagger})
\nonumber\\&+\Delta_1\sigma_{1z}+\Delta_2\sigma_{2z},
\end{align}
where $a^{\dagger}$ and $a$ are the single mode photon creation and
annihilation operators with frequency $\omega$, respectively, and
$\sigma_{i}\, (i=x,y,z)$ are the Pauli matrices. $2\Delta_1$,
$2\Delta_2$ are the transition frequencies of the two (discernible)
qubits. $g_{1}$ and $g_{2}$ are the qubit-photon coupling constants
for the two qubits respectively. (The fully symmetric case
corresponds to $g_1=g_2$, $\Delta_1=\Delta_2$.) First we make
unitary transformations to interchange $\sigma_x$ and $\sigma_z$ and
obtain $H^{\prime}_{tq}$. In Fock space, this Hamiltonian is
infinite dimensional with off-diagonal elements. However, by using
its $\mathbb{Z}_2$ symmetry with the transformation
$R^{\prime}=R\otimes \sigma_{1x}\otimes \sigma_{2x}$, where
$R=\exp(i\pi a^\dagger a)$, we obtain its solution analytically in
the Bargmann space \cite{br,barg}, in which the bosonic creation and
annihilation operators have the realizations
$a~\rightarrow~\frac{\partial}{\partial z}$,
$a^{\dagger}~\rightarrow~z$, with $z$ being a complex variable.

Setting $\omega=1$ and following the same procedure as \cite{pj}, we
obtain after a
Fulton-Gouterman transformation $U$ \cite{fg,pj},
$U^\dagger H^\prime_{tq} U=\left(
\begin{array}{cc}
H_+ &0 \\
0 & H_- \\
\end{array}
\right)$, where \begin{align}H_{\pm}=z\partial_z+g_{1}(z+\partial_
z)+g_{2}(z+\partial_
z)\sigma_{2z}\nonumber\\+\Delta_2\sigma_{2x}\pm\Delta_1
R\sigma_{2x},\label{+}\end{align} acting on the subspace where
$R^{\prime}$ has eigenvalues $\pm1$ respectively. We expand $H_\pm$
in the basis of $\{|e\rangle\otimes \varphi^\pm_1$,
$|g\rangle\otimes \varphi^\pm_2\}$, where $\varphi^\pm_1$ and
$\varphi^\pm_2$ are photon field wavefunctions. Making the
transformation $z\rightarrow-z$ to utilize the reflection symmetry
and denoting $\varphi_{3}(z)=\varphi_{1}(-z)$,
$\varphi_{4}(z)=\varphi_{2}(-z)$, we reduce the eigenvalue problem
to four coupled differential equations (see Eqs.
\eqref{z1}--\eqref{z4} in Appendix A). $\varphi^\pm_j$
($j=1,\ldots,4$) can be expanded into normalized extended coherent
states\begin{align} |n,-\alpha\rangle=\frac{e^{-\alpha^2/2+\alpha
z}}{\sqrt{n!}}(z-\alpha)^n\end{align} as
$\varphi^\pm_j=e^{\alpha^2/2}\sum_{n=0}^{\infty}\sqrt{n!}e^\pm_{j,n}|n,-\alpha\rangle$,
where $\alpha$ is a complex variable, with the recurrence relations
of $e^\pm_{j,n}$ as Eqs. \eqref{ac1}--\eqref{ac4} in Appendix A.

\begin{figure}[htbp]
\center
\includegraphics[bb=-40 0 1290
600,width=0.95\textwidth]{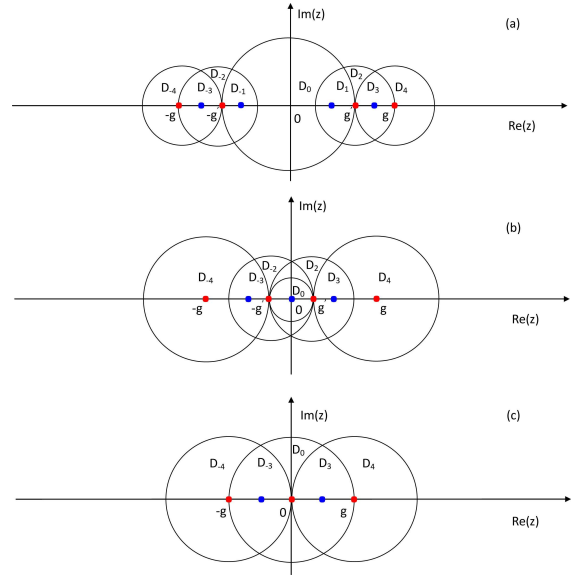}
\renewcommand\figurename{\textbf{FIG.}}
\caption[6]{(Color online) The singularity structure of $\varphi^\pm_j$ in the
complex plane, where $D_0$, $D_{\pm 2}$ and $D_{\pm4}$ denote the
circles of convergence of $\varphi^\pm_j$ expanded around $0$, $\pm
g^\prime$ and $\pm g$ respectively. Red/Light grey points denote
regular singular points, while blue/dark grey ones are ordinary
points. (a) $g^\prime\geq g/2$. $D_{\pm 1}=D_0\cap D_{\pm 2}$ and
$D_{\pm3}=D_{\pm 2}\cap D_{\pm 4}$ are the corresponding overlaps.
(b) $0<g^{\prime} \leq g/3$. $D_{\pm3}=D_{\pm 2}\cap D_{\pm 4}$. (c)
$g^\prime=0$. $D_{\pm3}=D_{0}\cap D_{\pm 4}$.\label{figure1}}
\end{figure}

Here the continued-fraction techniques, which work for the one-qubit
case, fail for $g_1\neq g_2$, because the equivalent recurrence
relation for $e^\pm_{1,n}$ has more than three terms if the
$e^\pm_{2,n}\ldots e^\pm_{4,n}$ are eliminated. However, we can
utilize the analyticity property of the Bargmann space
\cite{br,barg}. All the power series depend on four free initial
conditions $e^\pm_{j,0}$. Every expansion has a certain radius of
convergence $R_\alpha$, and we choose $|z-\alpha|<R_\alpha$ so that
the power series are absolutely convergent and a finite number of
terms suffices to compute the function reliably. In contrast to
numerical calculations in a truncated Hilbert space, where
convergence is found ``empirically'', convergence is a known
property in our scheme. Analyticity requires the wavefunctions in
the overlap of expansions around different points $\alpha$,
$\alpha^\prime$ to be the same, which furnishes four equations
$\varphi^\pm_j(z_0)=\varphi^{\prime\pm}_j(z_0)$ at a point $z_0$ in
the overlap \cite{bra2}. But there are eight free initial
conditions, so this will not impose enough constraints to obtain the
eigenvalue $E$ if $\alpha$, $\alpha^\prime$ are arbitrary ordinary
points of Eqs. \eqref{z1} -- \eqref{z4}. However, by analyzing the
structure of the recurrence relations for $e^\pm_{j,n}$ and
considering $\varphi^\pm_1(z)=\varphi^\pm_3(-z)$,
$\varphi^\pm_2(z)=\varphi^\pm_4(-z)$ at $z=0$, we find for some
special cases: $\alpha=\pm g$, $\pm g^\prime$, $0$, the free initial
conditions reduce to less than four. Choosing $\alpha=0$, $g^\prime$
and $g$, we denote $\varphi^\pm_j(z)$ by $\Phi^\pm_j(z)$,
$\phi^\pm_j(z)$ and $\psi^\pm_j(z)$, respectively. The corresponding
radii of convergence are
$R_{\alpha=0}=g^\prime$,~$R_{\alpha=g^\prime}=\min\{2g^{\prime},g-g^{\prime}\}$,~$R_{\alpha=g}=g-g^\prime$.
The singularity structure for $\varphi^\pm_j$ in the complex plane
is shown in Fig. \ref{figure1}.

\begin{figure}[htbp]
\center
\includegraphics[width=90mm]{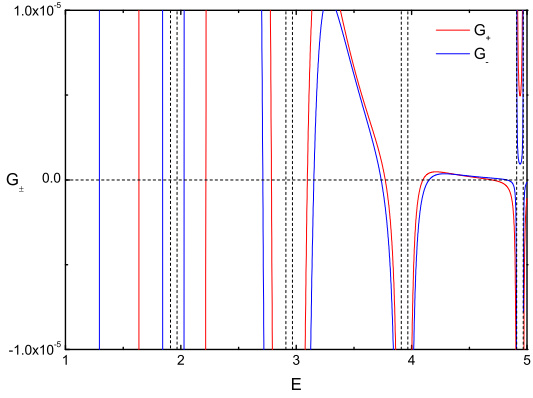}
\renewcommand\figurename{\textbf{FIG.}}
\caption[2]{The G-function of two-qubit Rabi model with
$\Delta_1=0.6$,~$\Delta_2=0.2$,~$\omega=1$,~$g=0.3$,~$g^\prime=0.18$.
Blue lines denote $G_-$ and red lines denote $G_+$, while dashed
vertical lines are base lines.\label{figure2}}
\end{figure}

Now there are eight free initial conditions
$\{e^{\pm,\alpha=0}_{1,0},$~$e^{\pm,\alpha=0}_{2,0}$,
$e^{\pm,\alpha=g^\prime}_{1,0}$,~$e^{\pm,\alpha=g^\prime}_{2,0}$,~$e^{\pm,\alpha=g^\prime}_{3,0}$,~$e^{\pm,\alpha=g}_{1,0}$,~
$e^{\pm,\alpha=g}_{2,0}$,~$e^{\pm,\alpha=g}_{4,0}\}$, determined by
the following eight equations
\begin{align}\label{z}
\Phi^{\pm}_j(z_0^\prime)-\phi^{\pm}_j(z_0^\prime)=0,~\phi^{\pm}_j(z_0)-\psi^{\pm}_j(z_0)=0,
\end{align}
where $z_0^\prime$ and $z_0$ are arbitrary points which satisfy
$|z-\alpha|<R_\alpha$, with which the analyticity condition of the
Bargmann space in the whole complex plane is fulfilled. The
condition for the linear equations (\ref{z}) to have non-trivial
solutions reads $G_\pm=\det (M_\pm)=0$ for the $8*8$ coefficient
matrix $M_\pm$ of Eq. \eqref{z} (see Eq. \eqref{de} in Appendix A),
which can be used to determine the eigenvalue $E$ and sequently the
eigenfunction $\varphi^{\pm}_j$ for $\alpha=0$, $g^\prime$, $g$. For
$g^\prime=0.6g$, we choose $z_0=(g^\prime+g)/2$ in $D_2$ and
$z^\prime_0=(g^\prime)^2/g$ in $D_1$, which satisfies
$z^\prime_0/g^\prime=(g^\prime-z^\prime_0)/(g-g^\prime)$ and obtain
$G_\pm$, shown in Fig. \ref{figure2}. The eigenenergy locates at the
zero of $G_\pm$, which does not vary with different $z_0$ and
$z_0^\prime$.
For $0<g^\prime<g/2$, as shown in Fig. \ref{figure1}(b), one can
choose $z_0^\prime=0$, then since
$\Phi_1(0)=\Phi_3(0)=e^{\prime\prime}_{1,0}$ and
$\Phi_2(0)=\Phi_4(0)=e^{\prime\prime}_{2,0}$, one finds
$\phi_j(0)=\Phi_j(0)$ is equivalent to $\phi_3 (0)=\phi_1(0)$,
$\phi_4(0)=\phi_2(0)$, and the matrix can be reduced to $6$
dimensions.

 For a special case $g^\prime=0$, i.e. $g_1=g_2$, we only
need to choose $\alpha=0$ and $g$ (see Fig. \ref{figure1}(c)), and
there are four equations
\begin{align} \Phi^{\pm}_j(z_0)=\psi^{\pm}_j(z_0) \end{align}
for four free initial conditions $\{e^{\pm,\alpha=0}_{1,0}$,
$e^{\pm,\alpha=g}_{1,0}$, $e^{\pm,\alpha=g}_{2,0}$,
$e^{\pm,\alpha=g}_{4,0}\}$. The lowest part of the spectra for four
sets of parameters are shown in Fig. \ref{figure2}, coinciding with
the numerical results very well. The total wave function can be
obtained as
\begin{align}
\chi^\pm=&e^{\alpha^2/2}\sum_{n=0}^{\infty}\sqrt{\frac{n!}{2}}[e^{\pm}_{1,n}\frac{1\pm
R}{2}|n,-\alpha\rangle(|e,e\rangle+|g,g\rangle)
\nonumber\\&+e^{\pm}_{1,n}\frac{1\mp
R}{2}|n,-\alpha\rangle(|e,g\rangle+|g,e\rangle)\nonumber\\&+
e^{\pm}_{2,n}\frac{1\pm
R}{2}|n,-\alpha\rangle(|e,e\rangle-|g,g\rangle)\nonumber\\&+
e^{\pm}_{2,n}\frac{1\mp
R}{2}|n,-\alpha\rangle(|g,e\rangle-|e,g\rangle)].
\end{align}
For $\alpha=0$, one has
\begin{align}
\chi^\pm=&\sum_n^{\infty}[e^{\pm}_{1,n}\sqrt{\frac{n!}{2}}(|n,e,e\rangle+|n,g,g\rangle)
\nonumber\\&+e^{\pm}_{2,n}\sqrt{\frac{n!}{2}}(|n,e,e\rangle-|n,g,g\rangle)\nonumber\\&+
e^{\pm}_{1,n\pm1}\sqrt{\frac{(n\pm1)!}{2}}(|n\pm1,g,e\rangle+|n\pm1,e,g\rangle)\nonumber\\&+
e^{\pm}_{2,n\pm1}\sqrt{\frac{(n\pm1)!}{2}}(|n\pm1,g,e\rangle-|n\pm1,e,g\rangle)],
\end{align}
where $n=0$, $2$, $4$, $\ldots$ for even parity and $n=1$, $3$, $5$,
$\ldots$ for odd parity. Now $\chi^\pm$ is a series of products of
Fock states and two-qubit Bell states. We can also obtain the
solution analytically with extended coherent state method
\cite{qing}, as shown in Appendix B.

\begin{figure}[htbp]
\center
\includegraphics[width=90mm]{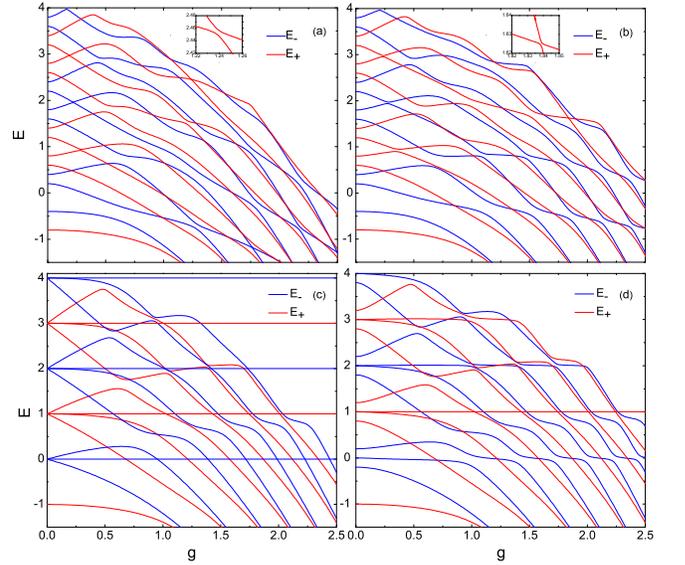}
\renewcommand\figurename{\textbf{FIG.}}
\caption[2]{(Color online) The spectra of two-qubit Rabi model with
(a) $\Delta_1=0.6$,~$\Delta_2=0.2$,~$\omega=1$,~$0\leq
g=g_1+g_2\leq2.5$,~$g_1=4g_2$.~ (b)
$\Delta_1=0.6$,~$\Delta_2=0.2$,~$\omega=1$,~$0\leq
g=g_1+g_2\leq2.5$,~$g_1=2g_2$.~(c)$\Delta_1=\Delta_2=0.5$,~$\omega=1$,~$0<g=g_1+g_2<2.5$,~$g_1=g_2$.
 ~(d)$\Delta_1=0.6$,~$\Delta_2=0.4$,~$\omega=1$,~$0\leq g\leq
2.5$,~$g_1=g_2$. Blue/Dark grey lines are eigenvalues with odd
parity, while red/light grew lines are eigenvalues with even
parity.\label{figure3}}
\end{figure}

\subsection{Exceptional solutions of the two-qubit quantum Rabi model}
From the recurrence relations of $e^\pm_{j,n}$ we find two kinds of
singular conditions for the eigenvalues at $E=n-(g^\prime)^2$ and
$E=n-g^2$, which can serve as two kinds of baselines and the second
one governs the asymptotics in the deep strong coupling regime of
the spectra. Exceptional solutions whose eigenvalues do not
correspond to zeros of $G_{\pm}$ occur at the baselines if the
parameters satisfy some conditions to lift the singularity of
$G_+(E)$ or $G_-(E)$. In this case, it may happen that the
recurrence relations for $e^\pm_{j,n}$, \eqref{ac1}--\eqref{ac4},
are cutoff at a certain $n$, and the eigenfunctions become
polynomials in $z$ if $\alpha=0$. However, not all exceptional
states posess this ``quasi-exact'' form. Unlike the Rabi model, the
conditions do not hold for $G_+(E)$ and $G_-(E)$ at the same time.
On the other hand, $G_+$ and $G_-$ are no longer closely related and
level crossings between states of different parity are thus not
confined to the baselines.

For identical couplings $g_1=g_2$ ($g^\prime=0$), the coupling term
is invariant under permutations of the qubits, which leads to a
special kind of exceptional eigenstate. By analyzing the recurrence
relations of $e^\pm_{j,n}$ for $\alpha=0$, we find polynomial
solutions at the baseline $E=N$, where $N$ is a nonnegative integer.
These states are very interesting for applications in quantum
information theory because they are essentially Fock states, without
coherent part as the quasi-exact eigenstates in the Rabi model and
the photon number is therefore strictly bounded from above. By
considering $\Phi_1(z)=\Phi_3(-z)$, $\Phi_2(z)=\Phi_4(-z)$ at $z=0$
and the recurrence relations of $e^\pm_{j,n}$ for $\alpha=0$, we
find $e^\pm_{1,n}=(-1)^ne^\pm_{3,n}$ and
$e^\pm_{2,n}=(-1)^ne^\pm_{4,n}$, and obtain a three term downward
recurrence relation for $e^\pm_{1,n}$ \begin{align}
e^\pm_{1,n}=&g^{-1}[(N-n-1)-\frac{(\Delta_2\mp(-1)^n\Delta_1)^2}{N-n-1}]e^\pm_{1,n+1}\nonumber\\&-(n+2)e^\pm_{1,n+2}
\label{la}
\end{align} with the initial conditions
\begin{align}
e^\pm_{1,N-1}=g^{-1}[\pm(-1)^{N-1}\Delta_1&-\Delta_2]e^\pm_{2,N}, \label{la1}\\
e^\pm_{1,N}=0\label{la2},
\end{align}
which satisfy $e^\pm_{j,n}=0$ for $n>N$. Then we can obtain
$e^\pm_{1,n}=f^\pm(n,N)e^\pm_{2,N}$ with non-zero $e^\pm_{2,N}$. But
$e^\pm_{1,-1}$ must vanish if no negative powers of $z$ appear, so
$f^\pm(-1,N)$ must equal to $0$. As seen from Eq. \eqref{la},
generally the condition for such states concerns $\Delta_1$,
$\Delta_2$ and $g$ if $\Delta_1\neq\Delta_2$, e.g., for $N=2$, the
condition reads \begin{align}
f^\pm(-1,2)=&[(2-\frac{(\Delta_1+\Delta_2)^2}{2})(1-(\Delta_1-\Delta_2)^2)-g^2]\nonumber\\&\times
g^{-3}(-\Delta_1-\Delta_2)=0.\label{ppp}
\end{align}This is the condition for an eigenstate with $E=2$ and
at most two photons. But it only exists if $\Delta_1$ and $\Delta_2$
are fine tuned with respect to the coupling $g$. This state is
therefore not so easy to prepare in an experiment, because the
coupling between the qubits and the radiation field are difficult to
control, whereas the qubit level splitting can be tuned very
precisely. If states exist where the condition $f^\pm(-1,N)=0$ does
{\it not} depend on $g$, but only on $\Delta_j$, they would be very
peculiar and interesting for applications.

Indeed, there exist two kinds of such states, with constant
eigenenergy for all coupling values. The first kind is the ``dark''
or ``trapping'' states \cite{rod-lara} if $\Delta_1=\Delta_2$ (see
Eq. \eqref{la1})
\begin{align}
|\psi\rangle_{ne}=\frac{1}{\sqrt{2}}|n,g,e\rangle-\frac{1}{\sqrt{2}}|n,e,g\rangle.
\end{align} In this
case, the Hamiltonian Eq.~\eqref{gq} possesses the full permutation
symmetry between the two spins and the Hilbert space separates
naturally into the singlet and the triplet part. The spin singlet
decouples from the radiation field, therefore the energy of all
singlet states does not depend on the coupling $g$ and the levels
cross those of the triplet states which interact with the radiation
field. The spectrum for
$\Delta_1=\Delta_2=0.5$,~$\omega=1$,~$0<g=g_1+g_2<2.5$, and
$g_1=g_2$ is shown in Fig. \ref{figure3}(c), where
$|\psi\rangle_{ne}$ correspond to the horizontal lines at $E=N$. The
qubits are in maximally entangled singlet Bell states, which are
robust even upon inclusion of dissipation \cite{rod-lara}. However,
this is entirely due to the full decoupling of the singlet states.
There is no entanglement between the qubits and the radiation mode.

Now, for the second case, it is quite surprising that a similar
state exists even when the full permutation invariance is partially
broken: If $g_1=g_2$ but $\Delta_1\neq\Delta_2$, only the coupling
term is invariant but the qubit energy is not. The function
$f^\pm(-1,1)$ reads
\begin{align}
f^\pm(-1,1)=g^{-2}(\pm\Delta_1-\Delta_2)[1-(\Delta_2\pm
\Delta_1)^2],
\end{align}
whose zeros are independent of $g$. So for even parity, if
$\Delta_1+\Delta_2=1$, there exists a state in the exceptional
spectrum with $E=1$, independent of $g$, although this state corresponds to strong coupling (and entanglement) between the qubits
and the radiation field as its
wavefunction
does explicitly depend on $g$,
\begin{align}
|\psi\rangle_e=\frac{1}{{\cal
N}}\left(\frac{2(\Delta_1-\Delta_2)}{g}|0,e,e\rangle-
 |1,e,g\rangle
+|1,g,e\rangle\right)\label{dk1}
\end{align}
with ${\cal N}=\sqrt{4(\Delta_1-\Delta_2)^2+2g^2}/g$. For odd
parity, if $\Delta_1-\Delta_2=1$ or $\Delta_2-\Delta_1=1$, there are
two corresponding eigenstates
\begin{align}
|\psi\rangle_{g1}=\frac{1}{{\cal
N}}\left(\frac{2(\Delta_1+\Delta_2)}{g}|0,e,g\rangle+
 |1,g,g\rangle
-|1,e,e\rangle\right),\label{dk2}\\
|\psi\rangle_{g2}=\frac{1}{{\cal
N}}\left(\frac{2(\Delta_1+\Delta_2)}{g}|0,g,e\rangle+
 |1,g,g\rangle
-|1,e,e\rangle\right),\label{dk3}
\end{align}
respectively,  where ${\cal
N}=\sqrt{4(\Delta_1+\Delta_2)^2+2g^2}/g$. The spectrum of the model
for $\Delta_1=0.6$, $\Delta_2=0.4$, $0<g=g_1+g_2<2.5$, and $g_1=g_2$
is shown in Fig.~\ref{figure3}(d). The exceptional solution
corresponds to the horizontal line of $E=1$, causing level crossings
within the same parity subspace, which was discovered numerically by
Chilingaryan and Rodr\'{i}guez-Lara \cite{2bite}. These states
contain at most one photon, so they are very interesting for
single-photon experiments. At the same time, they exist for all
coupling values, so they can be prepared without precise knowledge
of $g$. The qubit energies can be fine tuned to satisfy the
condition $\Delta_1+\Delta_2=\omega$. One can also obtain these
exceptional states in Fock space, as shown in Appendix C.

One may infer from Fig.~\ref{figure3}(a) and (b), that there are
level crossings in the spectrum between eigenstates with different
parity but not for the same parity, so that they can be labeled just
by two quantum numbers
--- energy level and parity, just as in the Rabi model (This, however, has
not yet been proven rigorously, see below). However, it has three
degrees of freedom and the dimension of the Hilbert space of the
discrete degrees of freedom is larger than the number of different
parity labels corresponding to the irreducible representations of
$\mathbb{Z}_2$. This renders the system non-integrable for general
values of model parameters \cite{br}, coinciding with what the
narrow avoided crossings in Fig. \ref{figure3}(a) and (b) indicate
\cite{xu}. Level crossings within a given parity subspace are caused
by an additional permutation symmetry in Fig.~\ref{figure3}(c) and
to the symmetry of the qubit-photon couplings in
Fig.~\ref{figure3}(d). Indeed, due to the possibility that
\eqref{de} may have a nullspace with dimension $>1$, level crossings
in the regular spectrum within a given parity subspace are not ruled
out in principle \cite{dicke}. One notes from Fig.~\ref{figure3}(d)
the appearance of horizontal baselines, because the singlet is
asymptotically decoupled for $g\rightarrow\infty$.

\section{Solutions of the generalized two-qubit Rabi models}\label{s3}
\begin{figure}[htbp] \center
\includegraphics[width=90mm]{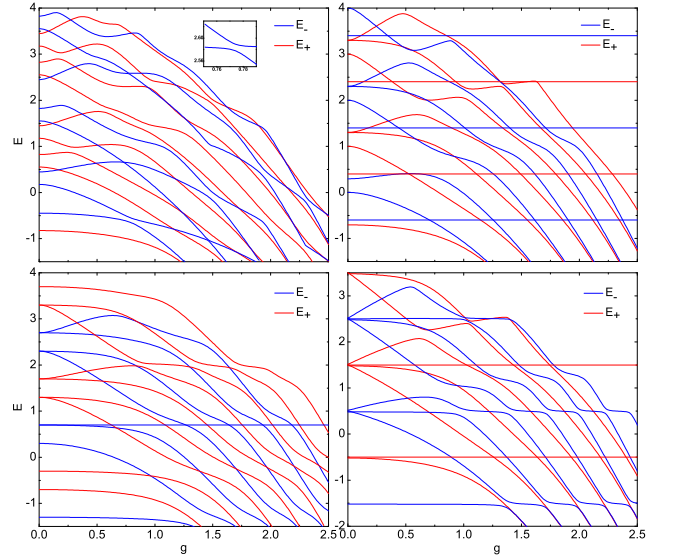}
\renewcommand\figurename{\textbf{FIG}}
\caption[2]{(Color online) The spectra of the generalized two-qubit
Rabi models with (a)
$\Delta_1=0.6$,~$\Delta_2=0.2$,~$\omega=1$,~$0<g=g_1+g_2<2.5$,~$g_1=4g_2$,~$J_x=0.2,~J_y=0.0,~J_z=0.0$.
(b)~$\Delta_1=0.5$,~$\Delta_2=0.5$,~$\omega=1$,~$0<g=g_1+g_2<2.5$,~$g_1=g_2$,~$J_x=0.1,~J_y=0.2,~J_z=0.3$.
(c)~$\Delta_1=0.1$,~$\Delta_2=0.7$,~$\omega=1$,~$0<g=g_1+g_2<2.5$,~$g_1=g_2$,~$J_x=0.7,~J_y=0.1,~J_z=0.3$.
(d)~$\Delta_1=0.6$,~$\Delta_2=0.4$,~$\omega=1$,~$0<g=g_1+g_2<2.5$,~$g_1=g_2$,~$J_x=0.5,~J_y=0.5,~J_z=0.5$.
Blue/Dark grey lines are eigenvalues with odd parity, while
red/light grew lines are eigenvalues with even
parity.\label{figure4}}
\end{figure}

Now we generalize the model by including
interactions between the two qubits which are commonly used to
generate entanglement \cite{gs,aa}, mandatory for applications in
quantum computation \cite{naga}. In our model, entanglement between
the qubits is produced naturally by the coupling to the radiation
field - nevertheless it is interesting to add a direct interaction
term, obtaining  more options to control the system. First we will
consider the anisotropic XYZ Heisenberg interaction
\begin{align}
H_{\text{XYZ}}=&\omega
a^+a+g_1(a+a^+)\sigma_{1x}+g_2(a+a^+)\sigma_{2x}\nonumber\\
&+\Delta_1\sigma_{1z}+\Delta_2\sigma_{2z}+\sum_{i=x,y,z}J_i\sigma_{1i}\sigma_{2i},
\end{align}
where $J_x$, $J_y$ and $J_z$ are the strength of XYZ Heisenberg
interaction in $x$, $y$, $z$ direction respectively.
$H_{\text{XYZ}}$ is $\mathbb{Z}_2$ invariant under the same
transformation as the two-qubit Rabi model. So using the same method
as above, we expand the photon field wave functions into the
normalized extended coherent state in the parity subspace as
$\varphi^{\pm}_j=\exp(\alpha^2/2)\sum_{n=0}^\infty
\sqrt{n!}f^{\pm}_{j,n}|n,-\alpha\rangle$ with the recurrence
relation of $f^{\pm}_{j,n}$ as Eqs. \eqref{ad1}--\eqref{ad4} in
Appendix D. Choosing $\alpha=0$, $g^\prime$ and $g$ for this model
as above, we can denote $\varphi^\pm_j(z)$ as $\psi^\pm_{j}(z)$,
$\phi^\pm_{j}(z)$ and $\Phi^\pm_{j}(z)$ respectively. Similarly,
there are eight equations
\begin{align}
\phi^\pm_j(z_0)=\psi^\pm_j(z_0),~
\phi^\pm_j(z^\prime_0)=\Phi^\pm_j(z^\prime_0)
\end{align}
for eight free initial conditions $\{f^{\pm,\alpha=0}_{1,0}$,
$f^{\pm,\alpha=0}_{2,0}$, $f^{\pm,\alpha=g^\prime}_{1,0}$,
$f^{\pm,\alpha=g^\prime}_{2,0}$, $f^{\pm,\alpha=g^\prime}_{3,0}$,
$f^{\pm,\alpha=g}_{1,0}$, $f^{\pm,\alpha=g}_{2,0}$,
$f^{\pm,\alpha=g}_{4,0}\}$, similar to Eq. \eqref{de}. The
determinant which is the function of eigenvalue $E$ must be zero if
non-trivial solutions to the equations exist. So the eigenvalues and
eigenstates are obtained. From Eqs. \eqref{ad3} and \eqref{ad4} we
find there are two kinds of baselines located at $E=n-g^2+J_x$ and
$E=n-(g^\prime)^2-J_x$. The first kind of baselines governs the
asymptotics for strong coupling. For XXX Heisenberg interaction and
dipole interaction, we just need to set $J_x=J_y=J_z$ and
$J_x=\epsilon$, $J_y=J_z=0$ respectively. The lowest part of the
spectra for four sets of parameters are shown in Fig. \ref{figure4}.
The narrow avoid crossing in Fig. \ref{figure4}(a) shows the
non-integrability of this model \cite{xu}, just like the two-qubit
Rabi model.

For the special case: $g^\prime=0$, i.e. $g_1=g_2$, we only need to
choose $\alpha=0$ and $g$, and there are four equations
\begin{align} \Phi^{\pm}_j(z_0)=\psi^{\pm}_j(z_0) \end{align}
for four free initial conditions $\{f^{\pm,\alpha=0}_{1,0}$,
$f^{\pm,\alpha=g}_{1,0}$, $f^{\pm,\alpha=g}_{2,0}$,
$f^{\pm,\alpha=g}_{4,0}\}$. Similar to the two-qubit Rabi model, we
find $f^\pm_{1,n}=(-1)^nf^\pm_{3,n}$ and
$f^\pm_{2,n}=(-1)^nf^\pm_{4,n}$. There are three kinds of base lines
located at $E=n-g^2+J_x$ and $E=n-J_x\pm J_y\pm J_z$. As usual,
exceptional solutions are located on the baselines if parameters
satisfy an additional constraint.

As in the non-interacting case, there exist
 exceptional solutions independent
of $g$ with constant eigenenergy. The condition just concerns the
qubits energy and qubit-qubit interaction strength. As discussed
above, by analyzing the recurrence relations of $f^\pm_{j,n}$ for
$\Phi^\pm_j(z)$ ($\alpha$, $g^\prime=0$), we find exceptional
solutions as polynomials in $z$ for $E=N-J_x\pm(-1)^N J_y\pm(-1)^N J_z$,
where $N$ is a nonnegative integer. For these states, the recurrence
relations of $f^\pm_{j,n}$ for $\alpha=0$ satisfy $f^\pm_{j,n}=0$
for $n>N$, and we find the three term downward recurrences for
$f^\pm_{1,n}$
\begin{align}
f^\pm_{1,n}=&g^{-1}[\frac{-(\Delta_2\pm(-1)^{n+1}\Delta_1)^2}{N-n-1\pm((-1)^N+(-1)^n)(J_y+J_z)}\nonumber\\
&\pm((-1)^N-(-1)^n)J_y\pm((-1)^N+(-1)^n)J_z\nonumber\\&+N-n-1-2J_x]f^\pm_{1,n+1}-
(n+2)f^\pm_{1,n+2} \label{lb} \end{align} with the initial condition
\begin{align}
f^\pm_{1,N-1}=g^{-1}[\pm(-1)^{N-1}\Delta_1&-\Delta_2]f^\pm_{2,N}, \label{lb0}\\
f^\pm_{1,N}=0,\label{lb1}
\end{align}
so that we can obtain $f^\pm_{1,n}=m^\pm(n,N)f^\pm_{2,N}$ with
non-zero $f^\pm_{2,N}$. But $f^\pm_{1,-1}$ must vanish if no
negative powers of $z$ appear, so $m^\pm(-1,N)=0$. As seen from Eq.
\eqref{lb}, this will give some constraint on $g$ in general, but
for $\Delta_1=\Delta_2$ (see Eq. \eqref{lb0}), the singlet Bell
state $|\psi\rangle_{n}$ is still the eigenstate with eigenenergy
$E=n-J_x-J_y-J_z$, shown in Fig. \ref{figure4}(b). They are robust
and survive the inclusion of XYZ Heisenberg interaction because
$\sum_{i=x,y,z}J_i\sigma_{1i}\sigma_{2i}|\psi\rangle_{n}=-(J_x+J_y+J_z)|\psi\rangle_{n}$.
However, they are not entangled with the radiation field. For $N=1$,
we obtain \begin{align} m^\pm(-1,1)=g^{-2}(\pm\Delta_1-\Delta_2)
[(J_x\pm J_y\pm 2J_z-1)^2\nonumber\\-(J_x\mp
J_y)^2-(\Delta_2\pm\Delta_1)^2]/(1-2J_y-2J_z)=0,\label{aq}
\end{align}
which is independent of $g$. So under this condition, there exists
an entangled eigenstate with even parity,
\begin{align}
|\psi\rangle_{e1}=&\frac{1}{\mathcal
{N}}[(1-2J_y-2J_z-\Delta_1-\Delta_2)|0,g,g\rangle\nonumber\\&+(1-2J_y-2J_z+\Delta_1+\Delta_2)|0,e,e\rangle\nonumber\\
&+\frac{(1-2J_y-2J_z)g}{\Delta_1-\Delta_2}(|1,g,e\rangle-|1,e,g\rangle)],\label{m1}
\end{align}with eigenenergy $E=1-J_x- J_y- J_z$ in the whole coupling regime, where $\frac{1}{\mathcal {N}}$ is a normalizing constant.
For odd parity, we just need to change the sign of $J_y$, $J_z$,
$\Delta_1$, and the spin direction of the first qubit in
$|\psi\rangle_{e1}$, corresponding to the horizontal line in Fig.
\ref{figure4}(c). If $J_x=J_y=J_z=0$, $|\psi\rangle_{e1}$ reduces to
$|\psi\rangle_{e}$. Even more interestingly, we find in the special
case $J_x\pm J_y\pm 2J_z=2$,
\begin{align}
m^\pm(-1,3)=0,
\end{align}
for the same parameters which yield $m^\pm(-1,1)=0$.
This corresponds to a second exceptional state with eigenenergy $E=3-J_x\mp
J_y\mp J_z$ in the spectrum for arbitrary coupling. This eigenstate reads
for even parity,
\begin{align}
|\psi\rangle_{e3}=\frac{1}{\mathcal
{M}}[(3-2J_y-2J_z-\Delta_1-\Delta_2)|0,g,g\rangle\nonumber\\+(3-2J_y-2J_z+\Delta_1+\Delta_2)|0,e,e\rangle\nonumber\\+\frac{(
2J_y+2J_z-3)(1-2J_y-2J_z-\Delta_2-\Delta_1)}{\sqrt{2}(1-2J_y-2J_z)}|2,g,g\rangle\nonumber\\+\frac{(
2J_y+2J_z-3)(1-2J_y-2J_z+\Delta_2+\Delta_1)}{\sqrt{2}(1-2J_y-2J_z)}|2,e,e\rangle\nonumber\\+\frac{\sqrt{6}(2J_y+2J_z-3)g}{2(\Delta_1-\Delta_2)}(
|3,g,e\rangle-|3,e,g\rangle)],\label{dk5}
\end{align}
where $\frac{1}{\mathcal {M}}$ is the normalizing constant. In this
case the spectrum contains two horizontal lines intersecting all
other levels, as shown in Fig. \ref{figure4}(d).

\section{Conclusion}\label{s4} We have studied the asymmetric two-qubit quantum Rabi model
and include (anisotropic) Heisenberg interactions between the
qubits. The spectra and eigenstates are obtained analytically using
Bargmann-space techniques. An equivalent alternative to this
solution of the two-qubit Rabi model is the method based on the
normalized extended coherent states \cite{qing} $|n,-\alpha\rangle$,
while continued-fraction techniques \cite{schweb,swa} are not
applicable. All models possess a parity symmetry, but this is not
sufficient for integrability because the discrete state space is
four-dimensional, whereas there are only two different parity
labels. We observe no level crossings in the regular spectrum within
the same parity chain (although they are not ruled out in principle
\cite{dicke}), and the narrow avoided crossings indicate thus the
non-integrability of the model, consistent with the criterion
proposed in \cite{br}. For special values of the qubit transition
frequencies, there exist exceptional solutions which cause level
crossings within the subspaces with fixed parity, reminiscent of the
quasi-exact solutions present in the Rabi model, but closely related
to the singlet states in the model with unbroken permutation
invariance. These simply structured eigenstates may be easily
prepared in current experimental set-ups because the condition for
their existence involves only the qubit energy splittings $\Delta_j$
and not the coupling $g$ to the radiation field. This makes them
especially well suited for applications in quantum storage and
transfer. The algebraic structure behind the possibility of this
novel type of exceptional eigenstate needs further clarification. We
may conjecture that the new class of states exists only for an even
number of qubits, where the permutationally invariant model
possesses a singlet sector.

\section*{Acknowledgements}
J.P. is thankful to Bo Zhou and Yibin Qian for helpful discussions.
D.B. thanks Karl-Heinz H\"ock for an important hint.
This work was supported by the National Natural Science Foundation
of China (Grants Nos 11035001, 10735010, 10975072, 11375086 and
11120101005), by the 973 National Major State Basic Research and
Development of China (Grants Nos 2010CB327803 and 2013CB834400), by
CAS Knowledge Innovation Project No. KJCX2-SW-N02, by Research Fund
of Doctoral Point (RFDP) Grant No. 20100091110028, by the Research
and Innovation Project for College Postgraduate of JiangSu Province
Grant Nos. CXLX13-025 and CXZZ13-0032, by the National Natural
Science Foundation of China (Grants No. 11347112), by the Project
Funded by the Priority Academic Program Development of Jiangsu
Higher Education Institutions (PAPD) and by the Deutsche
Forschungsgemeinschaft through TRR80.

\onecolumngrid
\appendix
\section{Solution of the
model without qubit interaction}\label{app1} First, we give some details about obtaining the
recurrence relations of $e^\pm_{j,n}$. In the Bargmann space, the
eigenvalue equations are reduced to four coupled differential
equations as
\begin{align}
(z+g)\frac{{\rm d}}{{\rm d}z}\varphi^\pm_{1}&=(E-gz)\varphi^\pm_{1}
\mp\Delta_1\varphi^\pm_{4}-\Delta_2\varphi^\pm_{2},
\label{z1}\\
(z+g^\prime)\frac{{\rm d}}{{\rm
d}z}\varphi^\pm_{2}&=(E-g^{\prime}z)\varphi^\pm_{2}
\mp\Delta_1\varphi^\pm_3-\Delta_2\varphi^\pm_1, \label{z2}\\
(z-g)\frac{{\rm d}}{{\rm d}z}\varphi^\pm_3&=(E+gz)\varphi^\pm_3
\mp\Delta_1\varphi^\pm_2-\Delta_2\varphi^\pm_4, \label{z3}\\
(z-g^\prime)\frac{{\rm d}}{{\rm
d}z}\varphi^\pm_4&=(E+g^{\prime}z)\varphi^\pm_4
\mp\Delta_1\varphi^\pm_1-\Delta_2\varphi^\pm_3, \label{z4}
\end{align}
where $g=g_1+g_2$, $g^\prime=g_1-g_2$. Expanding $\varphi^\pm_j$
($j=1$, $\ldots$, $4$) as
$\varphi^\pm_j=e^{\alpha^2/2}\sum_{n=0}^{\infty}\sqrt{n!}e^\pm_{j,n}|n,-\alpha\rangle$,
and substituting it into Eqs. \eqref{z1}--\eqref{z4}, we find the
recurrence relations for $e^\pm_{j,n}$
\begin{align}
(\alpha+g)(n+1)e^{\pm}_{1,n+1}&=(E-n-2\alpha
g-\alpha^2)e^{\pm}_{1,n}-(\alpha+g)e^{\pm}_{1,n-1}\mp\Delta_1e^{\pm}_{4,n}-\Delta_2e^{\pm}_{2,n},\label{ac1}\\
(\alpha+g^\prime)(n+1)e^{\pm}_{2,n+1}&=(E-n-2\alpha
g^\prime-\alpha^2)e^{\pm}_{2,n}-(\alpha+g^\prime)e^{\pm}_{2,n-1}\mp\Delta_1e^{\pm}_{3,n}-\Delta_2e^{\pm}_{1,n},\label{ac2}\\
(\alpha-g)(n+1)e^{\pm}_{3,n+1}&=(E-n+2\alpha
g-\alpha^2)e^{\pm}_{3,n}-(\alpha-g)e^{\pm}_{3,n-1}\mp\Delta_1e^{\pm}_{2,n}-\Delta_2e^{\pm}_{4,n},\label{ac3}\\
(\alpha-g^\prime)(n+1)e^{\pm}_{4,n+1}&=(E-n+2\alpha
g^\prime-\alpha^2)e^{\pm}_{4,n}-(\alpha-g^\prime)e^{\pm}_{4,n-1}\mp\Delta_1e^{\pm}_{1,n}-\Delta_2e^{\pm}_{3,n},\label{ac4}
\end{align}

Then we show the details of obtaining the eigenvalue $E$ with Eq.
\eqref{z}. To have a more convenient form for practical calculation,
we denote $\phi^\pm_j(z)=\sum_ke^\pm_{k,0}\phi_j^{k\pm}(z)$, $k=1$,
$2$, $3$, $\psi^\pm_j(z)=\sum_le^{\prime\pm}_{l,0}\psi_j^{l\pm}(z)$,
$l=1$, $2$, $4$, and
$\Phi_j(z)=\sum_me^{\prime\prime\pm}_{m,0}\Phi_j^{m\pm}(z)$, $m=1$,
$2$, as in \cite{dicke}, where for example, $\phi_j^{1\pm}(z)$ is
obtained by setting $e^\pm_{1,0}=1$ and $e^\pm_{2,0}$,
$e^\pm_{3,0}=0$ in linear equations \eqref{ac1}--\eqref{ac4}. Eqs.
\eqref{z} take the form of the following linear system
\begin{align}\label{de}
&\left(\begin{array}{cccccccc}
    \psi_1^{1\pm}(z_0) & \psi_1^{2\pm}(z_0) & \psi_1^{3\pm}(z_0) & -\phi_1^{1\pm}(z_0) & -\phi_1^{2\pm}(z_0) & -\phi_1^{3\pm}(z_0) &0 &0\\
    \psi_2^{1\pm}(z_0) & \psi_2^{2\pm}(z_0) & \psi_2^{3\pm}(z_0) & -\phi_2^{1\pm}(z_0) & -\phi_2^{2\pm}(z_0) & -\phi_2^{3\pm}(z_0) &0 &0\\
    \psi_3^{1\pm}(z_0) & \psi_3^{2\pm}(z_0) & \psi_3^{3\pm}(z_0) & -\phi_3^{1\pm}(z_0) & -\phi_3^{2\pm}(z_0) & -\phi_3^{3\pm}(z_0) &0 &0\\
    \psi_4^{1\pm}(z_0) & \psi_4^{2\pm}(z_0) & \psi_4^{3\pm}(z_0) & -\phi_4^{1\pm}(z_0) & -\phi_4^{2\pm}(z_0) & -\phi_4^{3\pm}(z_0) &0 &0\\
    0 & 0 & 0 & \phi_1^{1\pm}(z^\prime_0) & \phi_1^{2\pm}(z^\prime_0) & \phi_1^{3\pm}(z^\prime_0)
& -\Phi_1^{1\pm}(z^\prime_0)&-\Phi_1^{2\pm}(z^\prime_0) \\
    0 & 0 & 0 & \phi_2^{1\pm}(z^\prime_0) & \phi_2^{2\pm}(z^\prime_0) & \phi_2^{3\pm}(z^\prime_0)
& -\Phi_2^{1\pm}(z^\prime_0)&-\Phi_2^{2\pm}(z^\prime_0)\\
    0 & 0 & 0 & \phi_3^{1\pm}(z^\prime_0) & \phi_3^{2\pm}(z^\prime_0) & \phi_3^{3\pm}(z^\prime_0)
& -\Phi_3^{1\pm}(z^\prime_0)&-\Phi_3^{2\pm}(z^\prime_0)\\
    0 & 0 & 0 & \phi_4^{1\pm}(z^\prime_0) & \phi_4^{2\pm}(z^\prime_0) & \phi_4^{3\pm}(z^\prime_0)
& -\Phi_4^{1\pm}(z^\prime_0)&-\Phi_4^{2\pm}(z^\prime_0)
  \end{array}\right)%
\left(\begin{array}{c}
         e^\pm_{1,0} \\
          e^\pm_{2,0} \\
         e^\pm_{4,0} \\
         e^{\prime\pm}_{1,0} \\
        e^{\prime\pm}_{2,0} \\
         e^{\prime\pm}_{3,0}\\
e^{\prime\prime\pm}_{1,0}\\
e^{\prime\prime\pm}_{2,0}
       \end{array}\right)
=0.
\end{align}
For \eqref{de} to have a non-trivial solution, the determinant of
the above $8*8$ matrix must vanish, which determines the eigenvalue
$E$.

For the convergent powerseries, one chooses $z_0^\prime\in D_1$,
however, it is remarkable that if we choose $z_0^\prime \notin D_1$,
we can still obtain correct eigenvalues $E_n$. If we choose large
enough truncating order $n_{max}$, the zero of $G_{\pm}$ will
converge to the correct value even though the power series is not
convergent, because the wave functions are holomorphic in
$\mathbb{C}$ exactly at $E_n$, entailing a convergent power series
expansion at $z^\prime_0$ outside of $D_1$ \cite{braak}. However,
because the power series are not convergent, one will encounter very
large values for the $\phi_j(z_0^\prime)$, rendering it not as
convenient as the convergent one for small eigenenergies.
\onecolumngrid
\section{Solution obtained with extended coherent states method}
An alternative to the solution of the two-qubit Rabi model presented
here is the method based on normalized extended coherent states
\cite{qing} $|n,-\alpha\rangle$, which is the eigenstate of
$(a^\dag-\alpha)(a-\alpha)$. Defining
$A^\dagger=a^{\dagger}-\alpha$, we can rewrite the Hamiltonian
\eqref{+} in the positive parity space as
\begin{align}
H_+=&A^\dagger
A+\alpha^2+\alpha(A+A^\dagger)+\Delta_2\sigma_{2x}+\Delta_1\sigma_{2x}R+(A+A^\dagger+2\alpha)(g_1+g_2\sigma_{2z}).
\end{align}
We expand it in the diagonal presentation of $\sigma_{2z}$. Then
using the time independent Schr\"{o}dinger equation, making the
transformation $R=\exp(i\pi a^\dagger a)$ on it, and denoting
$\varphi_3=R\varphi_1$, $\varphi_4=R\varphi_2$, we obtain four
equations
\begin{align}
[A^\dagger A+\alpha^2+2\alpha
g+(g+\alpha)(A+A^\dagger)-E]\varphi_1+\Delta_2\varphi_2+\Delta_1\varphi_4=0,\label{2}\\
[A^\dagger A+\alpha^2+2\alpha
g^\prime+(g^\prime+\alpha)(A+A^\dagger)-E]\varphi_2+\Delta_2\varphi_1+\Delta_1\varphi_3=0,\label{3}\\
[A^\dagger A+\alpha^2-2\alpha
g-(g-\alpha)(A+A^\dagger)-E]\varphi_3+\Delta_2\varphi_4+\Delta_1\varphi_2=0,\label{4}\\
[A^\dagger A+\alpha^2-2\alpha
g^\prime-(g^\prime-\alpha)(A+A^\dagger)-E]\varphi_2+\Delta_2\varphi_3+\Delta_1\varphi_1=0,\label{5}
\end{align}
where $g=g_1+g_2$, $g^\prime=g_1-g_2$. Then we expand $\varphi_j$,
$j=1,2,3,4,$ in terms of the orthogonal extended coherent state as
$\varphi_j=e^{\alpha^2/2}\sum_{m=0}^{\infty}\sqrt{m!}e_{j,m}|m,-\alpha\rangle$,
and left multiply $|n,-\alpha\rangle$, we obtain the recursion
relations for $e_{j,n}$ which are the same as Eqs.
\eqref{ac1}--\eqref{ac4}.

In order to limit the number of free initial conditions, we choose
$\alpha=g,g^\prime,0$, then we obtain three expansions of the
wavefunction. They can be different only by a constant, which can be
chosen as 1, because the linearity of Eqs. \eqref{2}--\eqref{5}. For
$\alpha=g^\prime$, we have
$\phi_{j}=e^{(g^\prime)^2/2}\sum_{n=0}^{\infty}\sqrt{n!}a_{j,n}|n,-g^\prime\rangle$.
For $\alpha=g$, we have
$\psi_{j}=e^{g^2/2}\sum_{n=0}^{\infty}\sqrt{n!}\sum_{n=0}^{\infty}b_{j,n}|n,-g\rangle$
and for $\alpha=0$, we have
$\Psi_{j}=\sum_{n=0}^{\infty}\sqrt{n!}\sum_{n=0}^{\infty}c_{j,n}|n\rangle$.
If we left multiply the basic vector of the Bargmann space $\langle
0|e^{\beta a}$, we have
 \begin{align} \langle 0|e^{\beta
a}\varphi_{j}=\sum_{n=0}^{\infty}e_{j,n}\exp(\alpha\beta)(\beta-\alpha)^n.
\end{align}
As discussed in the Bargmann space, we have $8$ equations
\begin{align}
\langle 0|e^{\beta_1 a}\phi_j=\langle 0|e^{\beta_1 a}\psi_j,
~\langle 0|e^{\beta_2 a}\phi_j=\langle 0|e^{\beta_2 a}\Phi_j
\end{align}
for $8$ initial conditions
$\{b_{1,0},b_{2,0},b_{4,0},a_{1,0},a_{2,0},a_{3,0},c_{1,0},c_{2,0}\}$.
To have a convergent expansion series, we choose
$(\beta-\alpha)<R_\alpha$, where $R_\alpha$ is the convergent radius
of $e_{j,n}$. So according to the analysis in the Bargmann space, we
can choose $\beta_1=z_0$ and $\beta_2=z_0^\prime$ to obtain the some
eigenvalue and eigenstate as in the Bargmann space. We can also
choose
 $\alpha=-g,-g^\prime,0$ and the results will be the same.
\onecolumngrid
\section{$E=N$ exceptional solution in Fock space}
In this appendix, we try to obtain the $g$-independent exceptional
solution in Fock space. If for example, $M$ and $N$ are even, the
Hamiltonian in a closed odd parity basis of
$\{|M,e,g\rangle,|M,g,e\rangle,|M+1,g,
g\rangle,|M+1,e,e\rangle,\cdots,|N-1,g,g\rangle,|N-1,e,e\rangle,
|N,e,g\rangle,|N,g,e\rangle$ reads
\begin{align}
\left(
  \begin{array}{ccccccc}
   0 & 0&\sqrt{M-1}g_1 & \sqrt{M-1}g_2 & 0&0&\dots  \\
    0 & 0&\sqrt{M-1}g_2 & \sqrt{M-1}g_1 & 0&0&\dots  \\
   \sqrt{M-1}g_1&\sqrt{M-1}g_2& M+\Delta_1-\Delta_2 & 0 &\sqrt{M}g_1&\sqrt{M}g_2 &\dots \\
   \sqrt{M-1}g_2&\sqrt{M-1}g_1& 0&M+\Delta_2-\Delta_1 &\sqrt{M}g_2&\sqrt{M}g_1 &\dots \\
   \hdotsfor{7}
   \end{array}
   \right)\\
   \left(
    \begin{array}{ccccccc}
    \hdotsfor{7}\\
    \dots & \sqrt{N}g_1 & \sqrt{N}g_2 & N+\Delta_1-\Delta_2 & 0 & \sqrt{N+1}g_1 & \sqrt{N+1}g_2 \\
    \dots & \sqrt{N}g_2 & \sqrt{N}g_1 & 0 & N+\Delta_2-\Delta_1 & \sqrt{N+1}g_2& \sqrt{N+1}g_1 \\
   \dots& 0&0 & \sqrt{N+1}g_1 & \sqrt{N+1}g_2 & 0 & 0 \\
    \dots& 0&0  & \sqrt{N+1}g_2 & \sqrt{N+1}g_1 & 0 & 0
  \end{array}
\right).
\end{align}
To have a closed subspace, the coefficients of $|N+1,g,g\rangle$ and
$|N+1,e,e\rangle$ must be zero, so we have
\begin{align}
\sqrt{N+1}\large{g}_{\scriptstyle 1} \large{c}_{1,N}+\sqrt{N+1}g_2 c_{2,N}=0,\label{n1}\\
\sqrt{N+1}g_2 c_{1,N}+\sqrt{N+1}g_1 c_{2,N}=0, \label{n2}\end{align}
where $c_{1,N},c_{2,N}$ are the coefficients of
$|N,e,g\rangle,|N,g,e\rangle$ respectively. From Eqs. \eqref{n1} and
\eqref{n2} and $g_1$, $g_2>0$ we obtain $g_1=g_2$ and
$c_{1,N}=-c_{2,N}$. With the time-independent Schr\"{o}dinger
equation, we obtain
\begin{align}
\sqrt{N}g_1 c_{1,N-1} + \sqrt{N}g_2 c_{2,N-1} +
(N+\Delta_1-\Delta_2)c_{1,N}=Ec_{1,N},\\
\sqrt{N}g_2 c_{1,N-1} + \sqrt{N}g_1 c_{2,N-1} +
(N+\Delta_2-\Delta_1)c_{2,N}=Ec_{2,N}, \end{align} from which we can
obtain
\begin{align}E&=N\\
(\Delta_2-\Delta_1)c_{1,N}&=(\sqrt{N}g_1 c_{1,N-1} + \sqrt{N}g_2
c_{2,N-1}).
\end{align}
If $\Delta_1=\Delta_2$ and $c_{1,N-1}=c_{2,N-1}=0$, the eigenstate
becomes
$|\psi\rangle_{N}=\frac{1}{\sqrt{2}}(|N,g,e\rangle-|N,e,g\rangle)$,
the well known ``dark state'' or ``trapping state'' \cite{rod-lara}.
Else, in order to have a closed subspace, the coefficients of
$|M-1,g,g\rangle$ and $|M-1,e,e\rangle$ must be $0$, so we obtain
$E=M$ using the time-independent Schr\"{o}dinger equation as above,
which contradict the condition $E=N$. So, we can only choose $M=0$,
where $|M-1,e,e\rangle$ vanish automatically. There is a special
case: $M=0$ and $N=1$, then it is required $\Delta_1-\Delta_2=1=N$
or $\Delta_2-\Delta_1=1=N$, and the corresponding eigenstates are
$|\psi\rangle_{g1}$ (see Eq. \eqref{dk2}) and $|\psi\rangle_{g2}$
(see Eq. \eqref{dk3}) respectively. For even parity case, it is
required that $\Delta_1+\Delta_2=1$ or $-\Delta_1-\Delta_2=1$. The
second condition can not be satisfied, so we find only one
exceptional eigenstate $|\psi\rangle_{e}$ (see Eq. \eqref{dk1}).
\onecolumngrid
\section{The recurrence relations of $f^\pm_{j,n}$}\label{app2}
First we make unitary transformations
$S=\frac{1}{\sqrt{2}}(\sigma_{x}+\sigma_{z})$ to $H_{\text{XYZ}}$ to
interchange $\sigma_x$ and $\sigma_z$ and obtain
$H^\prime_{\text{XYZ}}$. Applying the same Fulton-Gouterman
transformation U \cite{pj,fg} as above, we obtain \begin{align}
 U^\dagger H^\prime_{\text{XYZ}}U= \left(
                              \begin{array}{cc}
                                H_{\text{XYZ}+} & 0 \\
                                0 & H_{\text{XYZ}-} \\
                              \end{array}
                            \right),
                            \end{align}
where $H_{\text{XYZ}\pm}=z\partial_z+g_1(z+\partial_
z)+g_2(z+\partial_z)\sigma_{2z}+\Delta_2\sigma_{2x} +J_x\sigma_{2z}
\pm\Delta_1R\sigma_{2x}\mp J_y R\sigma_{2z}\pm J_zR$ for two
invariant subspaces with eigenvalues of $R^\prime$ being $\pm 1$
respectively. We expand it in the diagonal representation of
$\sigma_{2z}$, denoting
$\varphi^\pm_{3,4}(z)=\varphi^\pm_{1,2}(-z)$, making the
transformation $z\rightarrow-z$ and obtain the time-independent
Schr\"{o}dinger equations
\begin{align}
(z+g)\frac{z}{dz}\varphi^\pm_{1}&=(E-J_x-gz)\varphi^\pm_{1}\mp(J_z-J_y)\varphi^\pm_3
-\Delta_2\varphi^\pm_{2}\mp\Delta_1\varphi^\pm_{4},\label{ae1}\\
(z+g^\prime)\frac{z}{dz}\varphi^\pm_{2}&=(E+J_x-g^\prime
z)\varphi^\pm_{2}\mp(J_z+J_y)\varphi_4
-\Delta_2\varphi^\pm_{1}\mp\Delta_1\varphi^\pm_{3},\label{ae2}\\
(z-g)\frac{z}{dz}\varphi^\pm_{3}&=(E-J_x+gz)\varphi^\pm_{3}\mp(J_z-J_y)\varphi^\pm_1
-\Delta_2\varphi^\pm_{4}\mp\Delta_1\varphi^\pm_{2},\label{ae3}\\
(z-g^\prime)\frac{z}{dz}\varphi^\pm_{4}&=(E+J_x+g^\prime
z)\varphi^\pm_{4}\mp(J_z+J_y)\varphi^\pm_2
-\Delta_2\varphi^\pm_{3}\mp\Delta_1\varphi^\pm_{1}.\label{ae4}
\end{align}
We expand the photon field wave functions into the normalized
extended coherent state in the parity subspace as
$\varphi^{\pm}_j=\exp(\alpha^2/2)\sum_{n=0}^\infty
\sqrt{n!}f^{\pm}_{j,n}|n,-\alpha\rangle$ and substitute them into
Eqs. \eqref{ae1}--\eqref{ae4} to obtain the recurrence relations for
$f^{\pm}_{j,n}$
\begin{align}
(n + 1)(g + \alpha)f^{\pm}_{1,n + 1}=&( E-n -\alpha^2- 2\alpha g-J_x
)f^{\pm}_{1,n}-(\alpha+g) f^{\pm}_{1,n -1}-\Delta_2f^{\pm}_{2,n}
\mp\Delta_1 f^{\pm}_{4,n}\mp(J_z-J_y)f^{\pm}_{3,n},\label{ad1}\\
 (n +1)(\alpha+g^{\prime})f^{\pm}_{2,n+ 1}=&(E-n -\alpha^2-2\alpha g^\prime+J_x
)f^{\pm}_{2,n}-(\alpha+g^{\prime})f^{\pm}_{2,n
-1}-\Delta_2f^{\pm}_{1,n}
\mp\Delta_1 f^{\pm}_{3,n}\mp(J_y+J_z)f^{\pm}_{4,n},\label{ad2}\\
 (n + 1)(\alpha -g)f^{\pm}_{3,n +1}=&(E-n -\alpha^2+2\alpha g-J_x )f^{\pm}_{3,n}- (\alpha-g ) f^\pm_{3,n - 1}-\Delta_2
f^{\pm}_{4,n} \mp\Delta_1 f^{\pm}_{2,n}\mp(J_z-J_y)f^\pm_{1,n},\label{ad3}\\
(n + 1)(\alpha -g^\prime)f^{\pm}_{4,n +1}=&(E-n-\alpha^2+2\alpha
g^\prime+J_x)f^{\pm}_{4,n}-(\alpha-g^\prime)f^{\pm}_{4,n-1}-\Delta_2
f^{\pm}_{3,n}\mp\Delta_1f^{\pm}_{1,n}\mp(J_y+J_z)f^{\pm}_{2,n}.\label{ad4}
\end{align}
which are then analyzed in a similar way as the $e^{\pm}_{j,n}$ in
Appendix \ref{app1}.

\twocolumngrid

\end{document}